\documentclass{osa-article}
\usepackage{siunitx}
\journal{osajournal}

\setprjcopyright

\articletype{Research Article}

\begin{document}

\title{Temporal contrast enhancement of energetic laser pulses by filtered SPM broadened spectra}

\author{Joachim Buldt\authormark{1,*} and Michael Müller\authormark{1} and Robert Klas\authormark{1,2} and Tino Eidam \authormark{3} and Jens Limpert\authormark{1,2,3,4} and Andreas Tünnermann\authormark{1,2,4}}

\address{\authormark{1}Institute of Applied Physics, Abbe Center of Photonics, Friedrich-Schiller-Universität Jena, Albert-Einstein-Str. 6, 07745 Jena, Germany\\
	\authormark{2}Helmholtz-Institute Jena, Fröbelstieg 3, 07743 Jena\\
	\authormark{3}Active Fiber Systems GmbH, Wildenbruchstr. 15, 07745 Jena, Germany\\
	\authormark{4}Fraunhofer Institute for Applied Optics and Precision Engineering, Albert-Einstein-Str. 7, 07745 Jena, Germany}

\email{\authormark{*}joachim.buldt@uni-jena.de} 



\begin{abstract}
We present a novel approach for temporal contrast enhancement of energetic laser pulses by filtered SPM broadened spectra. A measured temporal contrast enhancement by at least 7 orders of magnitude in a simple setup has been achieved. This technique is applicable to a wide range of laser parameters and poses a highly efficient alternative to existing contrast-enhancement methods.\\
\url{https://doi.org/10.1364/OL.42.003761}
\end{abstract}

With increasing pulse energy in ultrafast lasers, temporal contrast is an more and more important topic. In strong-field physics, where a high-energetic pulse interacts with matter, prepulses can already lead to ionization and modify the experimental conditions. It is therfore usually requiered, that the pulse contrast exceeds ten orders of magnitude. Common techniques to improve the temporal contrast are cross-polarized wave generation (XPW) \cite{Jullien:2005} and nonlinear ellipse rotation (NER) \cite{Homoelle:2002}. Even though these techniques are well established, XPW is limited to low pulse energies and average powers to avoid damaging the utilised crystal. The contrast enhancement achievable with NER is limited by the quality of the quarter-wave-plates \cite{Homoelle:2002}. For ultrafast lasers with short pulse-durations and  wide spectral ranges achromatic waveplates have to be used, but also these show a limit in the polarization preservation.

We present a novel approach that is simple to implement and almost peak-power maintaining with an efficiency in the order of 30\%, while being virtually unlimited in terms of contrast enhancement. The technique does not show any of the restricitions nonlinear ellipse rotation or cross-polarized wave generation show. We demonstrate the functionality of the method theoretically by a simulation and experimentally by a first proof-of-principle experiment.

\begin{figure*}[!h]
	\centering
	{\includegraphics[width=\linewidth, trim={0.cm 0.2cm 0 0.2cm}]{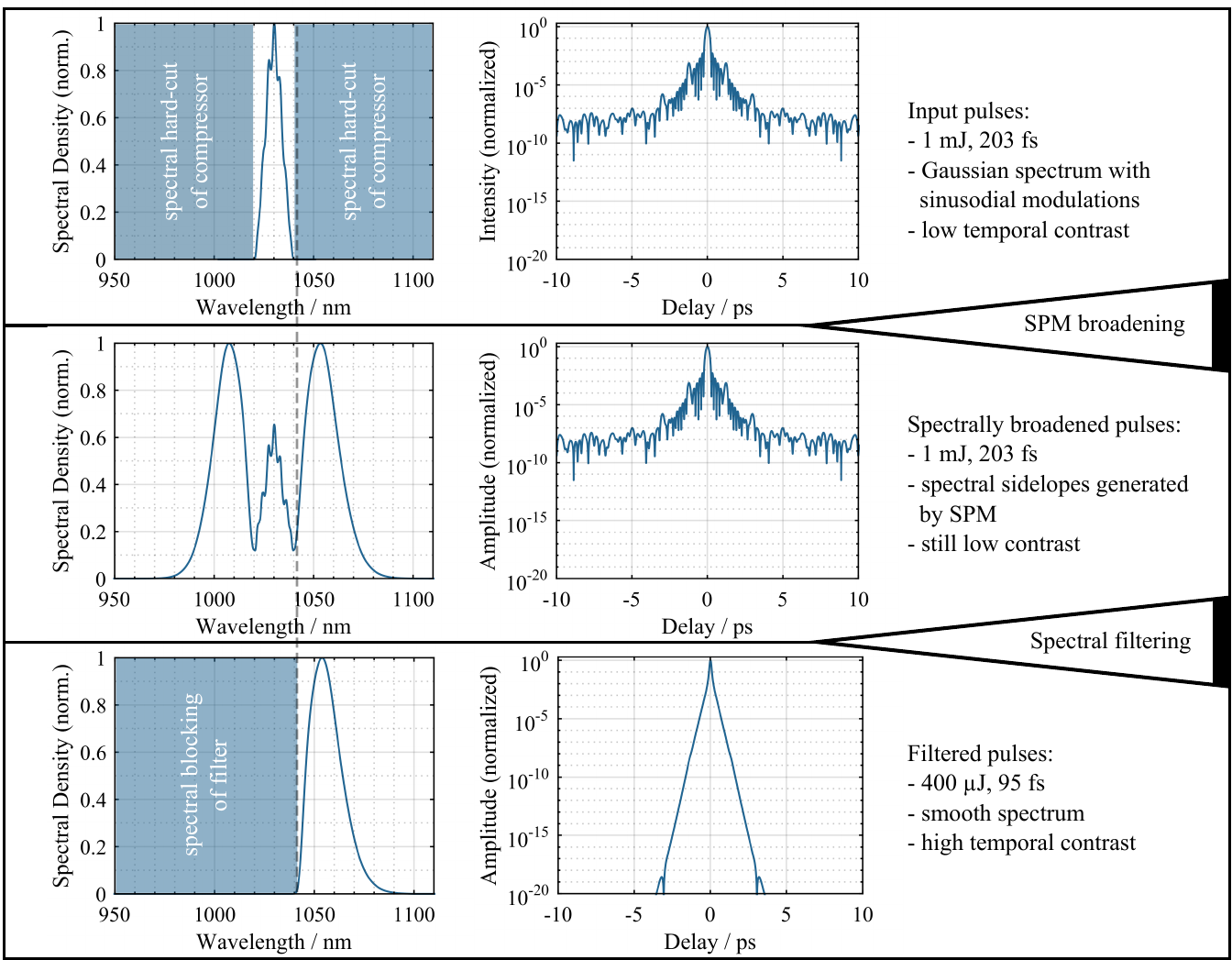}}
	\caption{Simulation of the SPM-based contrast-enhancement technique. For the simulation a Gaussian spectrum with sinusoidal modulations was used. The shaded area in the input spectrum indicates the spectral hardcut of the compressor. As one can see in the first row, the modulated gaussian spectrum generates a short pulse with strong pre- and post pulses in the timedomain. Due to self-phase-modulation the spectrum broadens, while the pulse shape and contrast remain unchanged. By applying a super-gaussian filter of the order of eight, only one sidelobe of the spectrum remains, which corresponds to a very short pulse with strongly improved contrast.}
	\label{fig:simulation}
\end{figure*}
\label{sec:examples}
The novel pulse-cleaning technique is based on self-phase-modulation (SPM) to broaden the spectrum of an incident low-contrast pulse. Since SPM is a peak power/intensity dependent effect, only the main pulse experiences significant spectral broadening and the lower intensity parts of the signal remain unaffected. The contrast improvement is achieved by suppressing the low-intensity parts of the signal with a spectral filter that only transmitts the wavelength-shifted part of the main pulse. Due to the spectral hard-cut in the compressor of a common chirped-pulse-amplification (CPA) system the intensity in the spectral region of the filtered light is initially equal to zero and thus, the achievable contrast is virtually unlimited.

A simulation has been carried out to demonstrate the principle of this pulse cleaning method. In Figure \ref{fig:simulation} spectral broadening due to SPM of pulses with central wavelength at 1030 nm has been numerically simulated. The spectrum of the pulses was generated as a gaussian spectrum with spectral modulations to introduce pre- and postpulses and a hardcut of a compressor at $ \SI{1020}{\nano\meter} $ and $ \SI{1042}{\nano\meter} $ has been included. The input pulses have a pulse energy of 1 mJ and a pulse duration of $ \SI{203}{\femto\second} $. With the numeric tool Fiberdesk \cite{Fiberdesk} the propagation of these pulses through a 0.6 m long hollow-core fiber with core-diameter of 250 µm and a nonlinearity of $ n_2 = \SI{9,8e-24}{\square\meter\per\watt} $, which corresponds  to a gas pressure of 1 bar argon \cite{Nisolo:1998}, has been simulated.  Afterwards a super-gaussian filter of the order of eight has been used to filter the spectrum so that there is no overlap with the incident spectrum confined by the compressors hardcut. The filtered spectrum contains approximately 40\% of the initial pulse energy and supports a pulse duration of $ \SI{95}{\femto\second} $, which shows that this method is peak-power preserving. It has to be mentioned here, that dispersion as well as the losses in the hollow-core fiber have been neglected in this simulation.

For a given laser only a suitable nonlinear medium has to be found to apply this contrast enhancement technique. For high average powers and mJ-class pulse energies hollow-core fibers already showed great performance \cite{Haedrich:2016}. For pulse energies in µJ regime, SPM in solid-core fibers can be used \cite{Jocher2012,Gaida:2015} and even towards high-energy joule-class laser pulses self-phase-modulation in bulk media is promising \cite{Lassone:2016} as well as SPM in multi-pass-cells \cite{Hanna:2017}. The presented technique is also applicable to a wide spectral range, i. e. only limited by availability of a transparent nonlinear medium. Additional, the spectral shift of the signal enables e. g. to use the short wavelength part of a broadened and contrast enhanced part of a Nd-Laser signal to seed an Yb-based system or vice versa. In the same way an ytterbium laser can be adapted to seed a titan-sapphire laser system with a high contrast and broadband signal. It is also possible to apply the technique twice on the same signal to further improve the contrast and preserve the central wavelength of the pulse.

\begin{figure}[!h]
	\centering
	{\includegraphics[width=0.5\linewidth]{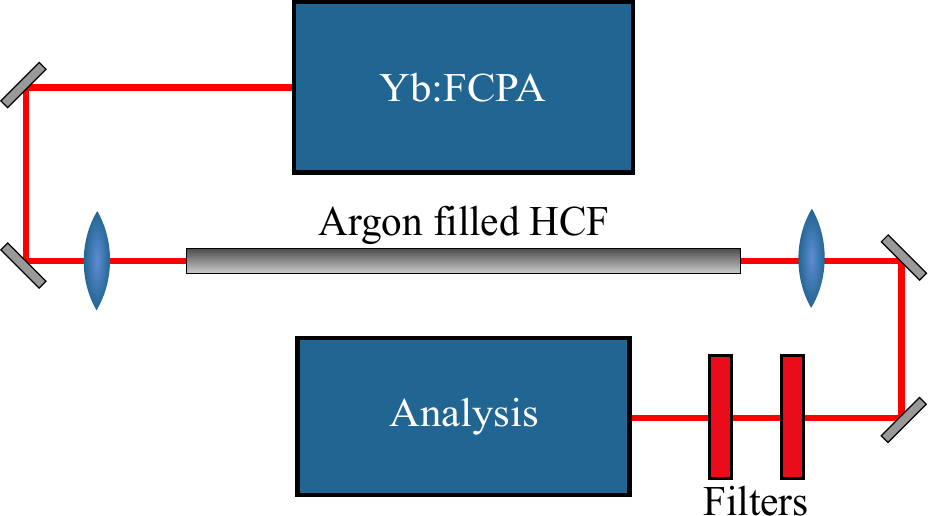}}
	\caption{Experimental setup used for the contrast enhancement. The light coming from the ultrafast fiber CPA system is coupled into a hollow-core fiber filled with argon. Afterwards it is spectrally filtered by two dielectric filters.}
	\label{fig:setup}
\end{figure}
\begin{figure}[!h]
	\centering
	{\includegraphics[width=0.5\linewidth, trim={0.cm 0.1cm 0 0.3cm}]{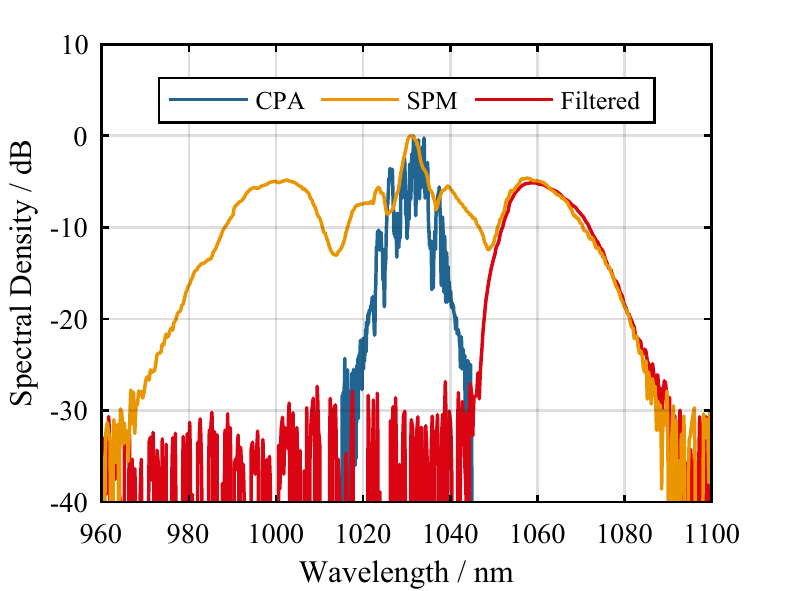}}
	\caption{Spectra of the incident, the SPM broadened and the filtered pulses. It is clearly visible, that the incident light from the Yb:FCPA system and the filtered light have no spectral overlap.}
	\label{fig:spectra}
\end{figure}

The fairly simple setup of our first proof-of-principle experiment is shown in Figure \ref{fig:setup}. A low-contrast signal from an ytterbim-based ultrafast fiber chirped-pulse-amplification system (Yb:FCPA) \cite{Kienel:2016,Mueller:2016} is transmitted through a noble-gas filled hollow-core fiber (HCF) and spectrally filtered afterwards. The filtered signal is then analysed by a third-order cross-correlator (Amplitude Technologies Sequoia) that offers a dynamic range of up to ten orders of magnitude. 
The used filters are dielectric long pass filters with the filter edge located at $ \SI{1050}{\nano\meter} $. Two identical filters with each a contrast of up to six orders of magnitude and a transmission of more then 95\% above $ \SI{1060}{\nano\meter} $ are utilized. 
In the experiment pulses from an ultrafast Yb:FCPA system delivering 1.9 mJ pulse energy, 290 fs pulse duration and a repetition rate of 1 kHz were used. 
The repetition rate in this experiment was limited by the acquisition speed of the cross-correlator.
The pulses emitted by the Yb:FCPA system are broadened in the HCF filled with argon at an absolute pressure of 1.2 bar. In Figure \ref{fig:spectra}, the spectra at different stages of the pulse cleaning are shown. The blue curve shows the spectrum of the Yb:FCPA pulses and is centered at $ \SI{1030}{\nano\meter} $ with a spectral hardcut of the compressor located at $ \SI{1042}{\nano\meter} $, i.e. virtually no light is emitted from the laser above $ \SI{1042}{\nano\meter} $. The Yb:FCPA pulses are spectrally broadened (yellow curve) and filtered (red curve). As one can see the filtered spectrum is centered at $ \SI{1060}{\nano\meter} $ and has an energy content of approximately 30\% of the HCF-output pulses. It is also visible, that it has no spectral overlap with the light of the Yb:FCPA system. 

\begin{figure}[!h]
	\centering
	{\includegraphics[width=0.5\linewidth, trim={0.cm 0.1cm 0 0.3cm}]{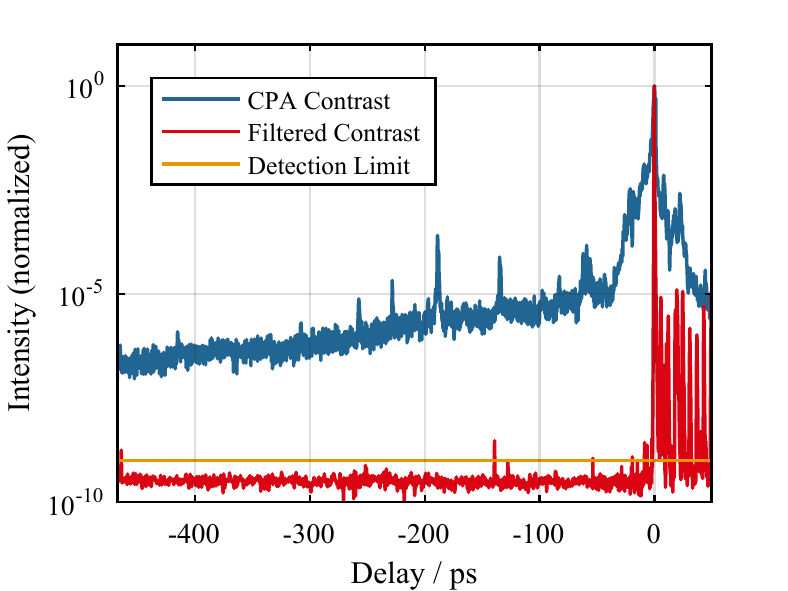}}
	\caption{Contrast measurement of the Yb:FCPA and the filtered pulses. The filtered pulses have a temporal contrast of at least $ 10^{-9} $ and the post-pulses originate from internal reflections in the filters and can be avoided by using wedged filters.}
	\label{fig:contrast}
\end{figure}

\begin{figure}[!h]
	\centering
	{\includegraphics[width=0.5\linewidth, trim={0.cm 0.1cm 0 0.3cm}]{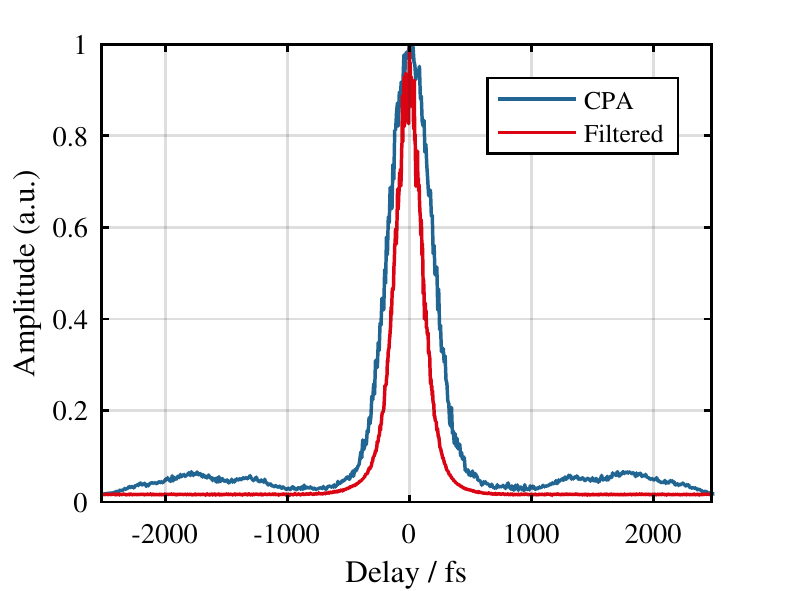}}
	\caption{Auto-correlation measurement of the pulses directly from the Yb:FCPA system and after the SPM-broadening and spectral filtering. The input pulses have a duration of $ \SI{290}{\femto\second} $ and are shortened to $ \SI{177}{\femto\second} $ by the filter.}
	\label{fig:ac}
\end{figure}

The contrast measurements are shown in Figure \ref{fig:contrast}, proving that the contrast of the filtered signal exceeds $ 10^{-9} $. Compared to the contrast of the incoming pulses, which is in the order of $ 10^{-2} $, we have achieved a contrast enhancement of at least seven orders of magnitude. Note that the measured contrast is limited by the dynamic range of the cross-correlator. Since we used two filters which provide six orders of magnitude contrast each, we can expect the contrast to be even higher. The post pulses visible in the filtered contrast measurement in Figure \ref{fig:contrast} originate from internal reflections in the spectral filters and can be avoided by using wedged filter substrates.
In Figure \ref{fig:ac} the auto-corellation measurement of the pulses is shown. The incident pulses have a duration of approximately $ \SI{290}{\femto\second} $ (FWHM) and the pre- and post pulses are also visible in the auto-correlation trace. The filtered pulses have a duration of $ \SI{177}{\femto\second} $ (FWHM) without any further compression after the transmission through the gas-filled hollow-core fiber.


\begin{figure}[!h]
	\centering
	{\includegraphics[width=0.5\linewidth, trim={0.cm 0.1cm 0 0.3cm}]{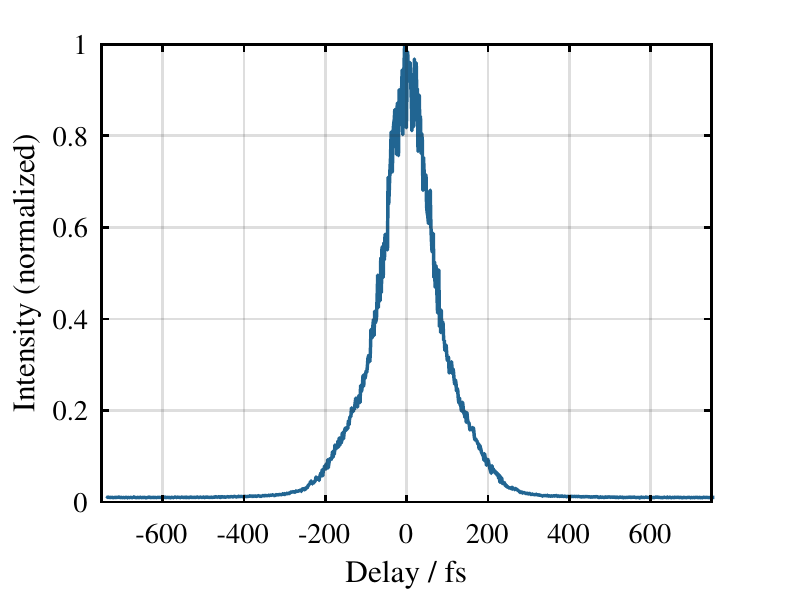}}
	\caption{Auto-correlation measurement of the filtered and compressed pulses. The pulse duration after a compressor with a GDD of $ \SI{-4200}{\square\femto\second} $ is $ \SI{95}{\femto\second} $.}
	\label{fig:ac_compressed}
\end{figure}

In an additional experiment a chirped-mirror-compressor with a group-delay-dispersion (GDD) of $ \SI{-4200}{\square\femto\second} $ has been included in the setup. In Figure \ref{fig:ac_compressed} an auto-correlation measurement of the compressed pulses is shown. These have a duration of $ \SI{95}{\femto\second} $ (FWHM) after the filter and compressor. Thus it is shown, that the spectrally broadened and filtered pulse can afterwards be compressed and the incident peak-power can in principle be restored.

In conclusion, we have demonstrated a novel technique for temporal contrast enhancement based on self-phase-modulation and spectral filtering. The method is simple to implement and allows highly efficient pulse cleaning with the contrast enhancement only being limited by the filter characteristics. Due to its simplicity and the scalability of SPM it is applicable for a wide range of different laser parameters, in terms of pulse duration, pulse energy and wavelength. The efficiency of the method can be further increased if the long- and short-wavelength part of the SPM-broadened spectrum are used.

\section*{Acknowledgement}
We acknowledge the help of our colleagues of the POLARIS group at the Institute for Optics and Quantum Electronics in Jena who kindly provided us with the Amplitude Technologies Sequoia.

\section*{Funding}
H2020 European Research Council (ERC) (617173, 670557).

\section{References}


\bibliography{sample}






\end{document}